\documentclass[12pt]{amsart}
\usepackage{amsmath,amssymb,tikz-cd}
\usepackage{hyperref}
\usepackage{mathtools}

\usepackage{xcolor}

\newcommand{\ndash}{\nobreakdash-\hspace{0pt}}
\newcommand{\Ndash}{\nobreakdash--}

\newcommand{\ii}{{\mathrm{i}}}
\newcommand{\dd}{{\mathrm{d}}} 
 
\newcommand{\EE}{{\mathrm{e}}}

\DeclareMathOperator{\Map}{Map}

\DeclareMathOperator{\ad}{ad}

\newtheorem{Thm}{Theorem}[section]

\newtheorem*{Thm*}{Theorem}
\newtheorem*{Lem*}{Lemma}

\theoremstyle{remark}

\theoremstyle{definition}

\newtheorem{rmk}[Thm]{Remark}

\newcommand{\bbZ}{{\mathbb{Z}}}

\newcommand{\de}{\partial}

\newcommand{\calM}{\mathcal{M}}

\newcommand{\sfdelta}{{\boldsymbol{\delta}}}
\newcommand{\sfdeltaeq}{{\boldsymbol{\delta}_{eq}}}

\newcommand{\targetq}{{\boldsymbol{q}}}

\def\gpd{\,\lower1pt\hbox{$\longrightarrow$}\hskip-.24in\raise2pt
               \hbox{$\longrightarrow$}\,}

\let\Hat=\widehat

\newcommand\qq{}
\newcommand\cmp[1]{{\qq Commun.\ Math.\ Phys.\ \bf #1}}

\newcommand{\calL}{{\mathcal{L}}}

\newcommand{\bea}{\begin{eqnarray}}
\newcommand{\eea}{\end{eqnarray}}

\def\g{\mathfrak g}

\def\Map{{\mathrm{Map}}}

\def\A{{\mathcal A}}
\def\Der{{\rm Der}}

\def\SDW{{S_{DW}}}
\def\SUV{{S_{UV}}}
\def\SIR{{S_{IR}}}
\def\MDW{{{\mathcal M}_{DW}}}
\def\MIR{{{\mathcal M}_{IR}}}
\def\MUV{{{\mathcal M}_{UV}}}
\def\R{{\mathbb R}}

\def\Icz{{\color{red}c_0}}
\def\Ico{{\color{red}c_1}}
\def\Ictwop{{\color{red}c_2^+}}
\def\Iphitwop{{\color{red}\phi_2^+}}
\def\Iphithree{{\color{red}\phi_3}}
\def\Iphif{{\color{red}\phi_4}}
\def\Uctwom{{\color{blue}c_2^-}}
\def\Ucthree{{\color{blue}c_3}}
\def\Ucf{{\color{blue}c_4}}
\def\Uphitwom{{\color{blue}\phi_2^-}}
\def\Uphio{{\color{blue}\phi_1}}
\def\Uphiz{{\color{blue}\phi_0}}

\def\beq{\begin{equation}\begin{aligned}}
\def\eeq{\end{aligned}\end{equation}}

\begin{document}

\title{Towards equivariant Yang-Mills theory}

\author{F.~Bonechi}\address{INFN Sezione di Firenze, Via G. Sansone 1, 50019 Sesto Fiorentino, 
Firenze, Italy}\email{francesco.bonechi@fi.infn.it}

\author[A.~S.~Cattaneo]{A.~S.~Cattaneo}
\address{Institut f\"ur Mathematik, Universit\"at Z\"urich\\
Winterthurerstrasse 190, CH-8057 Z\"urich, Switzerland}  
\email{cattaneo@math.uzh.ch}

\author{M.~Zabzine}\address{Department of Physics and Astronomy, Uppsala University, Box 516, 75120 Uppsala, Sweden}\email{maxim.zabzine@physics.uu.se}

\date{}

\maketitle

\begin{abstract}
We study four dimensional gauge theories in the context of an equivariant extension of the Batalin-Vilkovisky (BV) formalism. 
 We discuss the embedding of BV Yang-Mills (YM) theory into a larger  BV theory and their relation. Partial integration 
  in the equivariant BV setting (BV push-forward map) is performed explicitly for the abelian case. 
 As result,  we obtain a non-local homological  generalization of the Cartan 
 calculus and  a non-local extension of the abelian YM BV action 
  which satisfies the equivariant master equation. 
\end{abstract}

 \tableofcontents

\section{Introduction}

 During the last thirty years equivariant localization played an instrumental role in the derivation of many exact 
  results in quantum field theory (QFT), in particular in the context of supersymmetric gauge theories. In QFT  
   % 1the 
   equivariant localization is based on a clever combination of the supersymmetry transformations with BRST symmetry  in such 
    fashion that the resulting transformations are interpreted as the equivariant differential on the space of fields. 
     In our previous work \cite{equiv-BV} we proposed a framework for the treatment of such theories within 
      the equivariant extension of the Batalin-Vilkovisky (BV) formalism. In the present paper we continue to study this extension and  discuss the equivariant partial BV integration (equivariant BV push-forward) map. 
       Our main examples will be four-dimensional gauge theories. 
       
 We observe that the BV description of Yang-Mills (YM) theory given by Costello in \cite{Costello} can be obtained restricting 
  the BV action of Donaldson--Witten (DW) theory  (\cite{equiv-BV}, \cite{Ca-1}, \cite{CCFMRTZ}, \cite{Ik}) to a BV submanifold. 
   Both theories can be recast in the AKSZ formulation, provided that, in the case of YM, one allows in the source a {%2
   differential graded algebra (dga)} $\mathcal A $ that is not {%3 locally freely generated
   the external algebra of differential forms}.  
   The BV push-forward map from DW theory to YM theory corrects the Costello's BV YM action by terms that in the abelian case involve the zero modes of the  ultraviolet (UV) fields but in
   the non abelian case are packed in a full perturbative expansion in the YM fields. In the equivariant case, we perform the explicit 
      BV push-forward map in the case of abelian theory and obtain an effective action that is non-local and satisfies the equivariant master equation. From a geometrical point of view we can see this result as a non-local homological generalization of the Cartan calculus.  
 
 The paper is organized as follows: in Section \ref{s:BV-overview} we recall basic notions within the BV formalism 
  and the AKSZ construction of the solution of the classical master equation. We give the formal definition of  the BV push-forward 
   map and  list its formal properties. In Section \ref{s:equiv-BV-overview} we summarize the equivariant extension of 
    the BV formalism and the corresponding equivariant version of the AKSZ construction. We define formally the equivariant analog 
     of the BV push-forward map and discuss its properties. In Section \ref{s:DW} we define the 4d AKSZ theory to which we refer as the Donaldson--Witten (DW) theory. We discuss different formulations of this theory. In Section \ref{s:YM} we discuss
       the BV formulation of Yang--Mills (YM) theory and its relation to the DW theory. We briefly discuss the application of the 
        BV push-forward map to the abelian DW theory. In Section \ref{s:equiv-DW-YM} we introduce the equivariant extension of 
         DW AKSZ theory. For the case of abelian DW theory we describe explicitly the equivariant BV push-forward map.
          The final result is a non-local {%4 extension
          deformation} 
          of abelian YM theory which satisfies the equivariant master equation. This provides a non-local Cartan calculus on the dg algebra $\mathcal A$. 
         We summarize the paper and discuss the open questions in \ref{s:summary}. 
         Technical details regarding the Hodge decomposition and the relevant properties of the operators are collected in 
         appendix \ref{appendix_Hodge}
 
 \subsubsection*{Aknowledgements}

 We thank Pavel Mn\"ev for  discussions. A.S.C. acknowledges partial support of SNF Grant No. 200020\_192080 and of the Simons Collaboration on Global Categorical Symmetries.
  This research was (partly) supported by the NCCR SwissMAP, funded by the Swiss National Science Foundation.
 A.S.C. and  M.Z. thank INFN Sezione di Firenze where part of this work was carried out. F.B. and M.Z. want to thank ESI, Vienna, where part of this work was carried out. 
 
\bigskip
\bigskip
 
 \section{Overview of BV formalisms}\label{s:BV-overview}
 
 \subsection{BV}
 %{\color{red}to define some general notion from BV just to set the notations, describe AKSZ briefly}
The BV formalism was introduced in \cite{BV1, BV2, BV3} to study the (in)dependence of the partition function (or, more generally, expectation values of observables) from the gauge fixing. In this setting, the space of fields is extended to an odd symplectic manifold (for bookkeeping, it is convenient to have an additional $\bbZ$\ndash grading with the symplectic form having degree $-1$; in this paper, we assume that the parity is the modulo two reduction of the integer degree). The action is extended to a BV action $S$ (of degree $0$) satisfying the quantum master equation
\begin{equation}\label{QME}
\frac12\{S,S\}-\ii\hbar\Delta S=0.
\end{equation}
Here $\{\ ,\ \}$ denotes the BV bracket, i.e., the odd Poisson bracket defined by the odd symplectic structure, and $\Delta$ is the BV Laplacian. The latter is a second-order differential operator which in Darboux coordinates $p_1,\dots,p_n,q^1,\dots,q^n$ (the parity of each $p$ being opposite to that of a the corresponding $q$) with respect to the standard Berezinian density $\mu_\text{stand}=\dd^nq\,\dd^np$ takes the form $\Delta=\frac\de{\de p_i}\frac\de{\de q^i}$. 

The first theorem of the BV formalism states that, if $f=\Delta g$ and $\calL$ is a Lagrangian submanifold on which the integral of $g$ converges, then
\[
\int\limits_\calL f \mu_\text{stand}^{\frac12}= 0.
\]
Here $\mu_\text{stand}^{\frac12}$ denotes the half-density associated to $\mu_\text{stand}$. Its restriction to a Lagrangian submanifold is canonically a density.

This fact implies the central theorem of the BV formalism which states that, for a family $\calL_t$ of Lagrangian submanifolds smoothly depending on $t$,
\[
I_t := \int\limits_{\calL_t} f \mu_\text{stand}^{\frac12}
\]
is constant, under the assumptions that $\Delta f=0$ and that for each $t$ the integral converges. In fact, one can show that
\[
\frac\dd{\dd t} I_t = \int\limits_{\calL_t} (\Delta (\dot\psi f)+\dot\psi\Delta f) \mu_\text{stand}^{\frac12}
= \int\limits_{\calL_t} \dot\psi\Delta f \mu_\text{stand}^{\frac12},
\]
where we have used the previous theorem. Here $\dot\psi$ describes the variation of the gauge fixing. If $\Delta f= 0$, we get the result. For applications in physics, one considers $f=\EE^{\frac\ii\hbar S}$, and $S$ is called the BV action.

The application in field theory then consists  in replacing the ill-defined integral on the reference Lagrangian $\calL_0$, which naively describes the original divergent integral times a vanishing integral over the ghosts, with a well-defined integral over some nearby $\calL_t$ (gauge fixing).

The formalism can actually be extended globally to an odd symplectic manifold $\calM$. In this case, the definition of the BV Laplacian requires the choice of a Berezinian density $\mu$ such that the resulting differential operator squares to zero. Alternatively, one can observe \cite{Khudaverdian:1989si, K-2} that there is a canonical Laplacian $\Delta$ acting on half-densities. For every half-density $\sigma$ of the form $\Delta\tau$ and every Lagrangian submanifold $\calL$ on which $\tau$ is integrable, one has
\begin{equation}\label{e:Deltazero}
\int\limits_\calL \sigma = 0.
\end{equation}
Moreover, for every half-density $\sigma$ satisfying $\Delta\sigma=0$, one has that
\[
I_t := \int\limits_{\calL_t} \sigma
\]
is constant, under the assumption that for each $t$ the integral converges. Typically, one chooses $\sigma$ of the form
$\EE^{\frac\ii\hbar S}\sigma_0$ with $\sigma_0$ a reference half-density satisfying $\Delta\sigma_0=0$. More generally, we have
\begin{equation}\label{e:dotI}
\frac\dd{\dd t} I_t =  \int\limits_{\calL_t} \dot\psi\,\Delta\sigma,
\end{equation}
where $\dot\psi$ describes the variation of the gauge fixing.

In field theory, all this has to be regularized, as we are in an infinite-dimensional context, so neither the integral against $\mu_\text{stand}^{\frac12}$ nor the BV Laplacian are defined. For the considerations in the present paper, it is enough to ignore the BV Laplacian, i.e., to assume that $S$ satisfies the classical master equation
\begin{equation}\label{CME}
\{S,S\}=0.
\end{equation}
The integral is on the other hand understood in perturbation theory around a nondegenerate Gaussian (the choice of gauge fixing achieves this goal).

\subsection{AKSZ}
The AKSZ formalism \cite{AKSZ:geometry_of_BV} is a general procedure to construct solutions of the classical master equation (\ref{CME}) (one checks a posteriori, by restriction to the variables of degree $0$ what  physical theory it describes).

The construction, in $d$ dimensions, requires an odd symplectic manifold $Y$, with exact symplectic form $\omega_Y=\dd\theta_Y$ of degree $d-1$, endowed with a solution $S_Y$ (of degree $d$) of the associated classical master equation. Given a closed $d$\ndash manifold $\Sigma_d$ (our space--time), to these data one associates the odd symplectic manifold $\calM=\Map(T[1]\Sigma_d,Y)$ and a BV action $S$ satisfying the classical master equation. 

We present the construction explictly for $Y$ a finite-dimensional vector space and $\theta_Y=p_i\,\dd q^i$. In this case, $\calM$ has Darboux coordinates determined by the superfields ${\mathbf P}_i$s and ${\mathbf Q}^i$s, each of which is a sum of differential forms on $\Sigma_d$ of all degrees: if a superfield corresponds to a target variable of degree $n$, then its $k$-form component  is assigned degree $n-k$. % (and parity the modulo two reduction of this integer). 
The odd symplectic form on $\calM$ is then defined by 
\begin{equation}
\omega=\int\limits_{\Sigma_d} \delta {\mathbf P}_i \wedge \delta {\mathbf Q}^i~,
\end{equation}
 where in each summand the integral selects the $d$\ndash form. The BV action is defined as
\begin{equation}\label{e:SAKSZ}
S = \int\limits_{\Sigma_d} ({\mathbf P}_i\dd {\mathbf Q}^i + S_Y({\mathbf P},{\mathbf Q})).
\end{equation}

\begin{rmk}\label{r:ggf}
 A typical gauge fixing in an AKSZ theory is obtained as follows. One picks a Riemannian metric $g$ and defines the codifferential $\dd^\dagger$ (see Appendix~\ref{appendix_Hodge}). Then one defines $\calL_g$ as the Lagrangian submanifold obtained by requiring each differential form in the superfields to be in the image of $\dd^\dagger$.
\end{rmk}
 
 \subsection{BV push-forward}
 %{ \color{red}{define formally BV push-forward map and list its formal properties}}
The central theorem of the BV formalism can be extended to partial integration (BV pushforward).\footnote{We follow the terminology of \cite{CattaneoMnevReshetikhin2018} and refer to its Section~2.2 for more details. To the best of our knowledge, the BV pushforward was first used in \cite{CattaneoRossi2001}. It was then realized by A. Losev that it can be used to define Wilsonian renormalization in a BV compatible way; this was first put to use, in the case of $BF$ theories, in P. Mn\"ev's thesis \cite{Mnev2008}. It was then extensively used by K. Costello \cite{Costello} to study the renormalization of gauge theories.}

Namely, given a product $\calM = \calM_{IR} \times\calM_{UV}$ of odd symplectic manifolds to which we refer as UV and IR respectively  
and with total BV Laplacian $\Delta=\Delta_{IR}+\Delta_{UV}$, one has, for every half-density $\sigma$ on $\calM$ satisfying $\Delta\sigma=0$ and for a family $\calL_{UV,t}$ of Lagrangian submanifolds of $\calM_{UV}$, that
\[
\sigma_{IR}(t) := \int\limits_{\calL_{UV,t}} \sigma 
\]
defines a family of half-densities $\sigma_{IR}(t)$ on $\calM_{IR}$ satisfying $\Delta_{IR}\sigma_{IR}=0$ for every $t$. Moreover, $
\frac\dd{\dd t}\sigma_{IR}$ is $\Delta_{IR}$\ndash exact; in view of \eqref{e:Deltazero}, this change of $\sigma_{IR}$ is irrelevant in the BV formalism.

 These results are obtained from the BV theorems we have listed above. In fact,
\[
\Delta_{IR} \sigma_{IR}(t) = \int\limits_{\calL_{UV,t}} \Delta_{IR} \sigma =  \int\limits_{\calL_{UV,t}} \Delta\sigma -  \int\limits_{\calL_{UV,t}} \Delta_{UV}\sigma 
=  -\int\limits_{\calL_{UV,t}} \Delta_{UV}\sigma
= 0,
\]
where we have used \eqref{e:Deltazero} on $\calM_{UV}$.
Similarly, by \eqref{e:dotI} on $\calM_{UV}$,
\[
\frac\dd{\dd t}\sigma_{IR}(t) := \int\limits_{\calL_{UV,t}} \dot\psi\,\Delta_{UV} \sigma .
\]
Therefore,
\begin{equation}\label{e:dotIpf}
\frac\dd{\dd t}\sigma_{IR} (t) =  \int\limits_{\calL_{UV,t}} \dot\psi\,\Delta_{UV} \sigma 
= \int\limits_{\calL_{UV,t}} \dot\psi\,\Delta\sigma - \Delta_{IR}\int\limits_{\calL_{UV,t}} \dot\psi\,\sigma,
\end{equation}
which yields the result if $\Delta\sigma=0$.

Typically, we have reference half-densities $\sigma_0$ on $\calM$ and $\sigma_{0, IR}$ on $\calM_{IR}$, and $\sigma=\EE^{\frac\ii\hbar S}\sigma_0$. In this case, we write $\sigma_{IR}=\EE^{\frac\ii\hbar S_\text{eff}}\sigma_{0,IR}$ and  call $S_\text{eff}$ the effective action. 

For applications in field theory (in particular, in this paper), the odd symplectic manifolds are affine spaces (with a given choice of global Darboux coordinates), and the reference half-densities are the standard ones. The effective action is computed in perturbation theory around nondegenerate Gaussians.

  \section{Overview of equivariant  BV formalisms}\label{s:equiv-BV-overview}
 
 \subsection{Equivariant BV}\label{s:equiBV}
 In \cite{equiv-BV} an equivariant version of the BV formalism was introduced. The first observation was that, if the quantum master equation is violated,
 \begin{equation}\label{e:Delta=T}
 \frac12\{S,S\}-\ii\hbar\Delta S=:T\not=0,
 \end{equation} 
 then one still has the central theorem for BV integrals and BV pushforwards if one restricts oneself to integrating on Lagrangian submanifolds on which $T$ vanishes (we call them $T$\ndash Lagrangian submanifolds); see Section \ref{eq_BV_push_fwd}.
 
 In particular, the case of the extension of an AKSZ theory to the Cartan model for equivariant cohomology was considered. Namely, one starts with an infinitesimal action of a Lie algebra on the space--time $\Sigma_d$. We denote by $\{v_a\}$ the vector fields corresponding to some basis $(e_a)$ and introduce degree\ndash two variables $u^a$. One can then define the modified action
 \begin{equation}\label{e:Sc}
 S_{eq} = S + u^a S_{\Hat \iota_{v_a}},
 \end{equation}
 where $S$ is as in \eqref{e:SAKSZ}, and
 \begin{equation}\label{e:Siota}
 S_{\Hat \iota_{v_a}} =  \int\limits_{\Sigma_d} {\mathbf P}_i\,\iota_{v_a} {\mathbf Q}^i
 \end{equation}
 with $\iota_{v_a}$ denoting the contraction operator and $\Hat\iota_{v_a}$ its extension to the space of fields. 
 It then follows that
 \begin{equation}\label{e:ScSc}
 \frac12\{S_{eq},S_{eq}\} +u^a S_{\Hat L_{v_a}} = 0,
 \end{equation}
 where
 \begin{equation}\label{e:SL}
 S_{\Hat L_{v_a}} =  \int\limits_{\Sigma_d} {\mathbf P}_i\,L_{v_a} {\mathbf Q}^i
 \end{equation}
 with $L_{v_a}$ denoting the Lie derivative with respect to $v_a$ and $\Hat L_{v_a}$ its extension to the space of fields. Assuming $\Delta S =0$ and $\Delta S_{\Hat \iota_{v_a}}=0$,\footnote{Both conditions are reasonable, as, formally, $\Delta$ on these functionals corresponds to taking the trace on $\Omega^\bullet(M)$ of operators of degree different from zero.} we get
 \begin{equation}\label{e:Tequiv}
 T = - u^a S_{\Hat L_{v_a}}.
 \end{equation}
 It follows that $\calL_g$ as in Remark~\ref{r:ggf} is a $T$\ndash Lagrangian submanifold if the metric $g$ is invariant.
 
 In this paper, we consider a variation on this theme. Namely, we consider a BV theory (specifically, abelian Yang--Mills in Costello's BV formulation{, see Chapter 6 in \cite{Costello}}) which is based on a generalized AKSZ construction. Since the source algebra admits Lie derivatives $L_{v_a}$ and so {%6  their lift
 has lifts $\Hat{L}_{v_a}$ to the space of fields, we look for an action
  which satisfies \eqref{e:ScSc} with $S_{\Hat L_{v_a}}$ as in \eqref{e:SL}. However since there is not a standard Cartan calculus in the source and in particular no contraction vector field $\iota_{v_a}$, we cannot use the solution as in \eqref{e:Sc}. We will rather obtain a solution to \eqref{e:ScSc} that has essentially the same properties as the one above; in particular, $\calL_g$ is a $T$\ndash Lagrangian submanifold for every invariant metric $g$. On the other hand, we will obtain a generalized Cartan calculus that includes a new definition $\{S_{\Hat \iota_{v_a}},\ \}$ of the contraction operator together with higher operations, in a way that is compatible with the theory at hand.

% {\color{red}to define equivariant BV just to set the notation and some general ideas,  maybe mention equivariant AKSZ briefly}

 \subsection{Equivariant BV push-forward}\label{eq_BV_push_fwd}
 Suppose $S$ satisfies \eqref{e:Delta=T}, i.e.,
 \[
 \Delta f = \left(\frac\ii\hbar\right)^2  T\, f
 \]
 with $f= \EE^{\frac\ii\hbar S}$. Consider the half-density $\sigma = f\sigma_0$, with $\sigma_0$ a reference $\Delta$\ndash closed half-density. We then have
 \[
 \Delta\sigma = \left(\frac\ii\hbar\right)^2 T\,\sigma.
 \]
  We now perform the BV pushforward with the extra assumption that $T=T_{IR}+T_{UV}$ with $T_{IR}$ and $T_{UV}$ functions on $\calM_{IR}$ and ${\calM}_{UV}$ respectively. 
  The half-density $\sigma_{IR}(t)$ on $\calM_{IR}$ defined by the pushforward along a $T_{UV}$-Lagrangian $\calL_{UV,t}$ on $\calM_{UV}$ then satisfies
 \[
 \Delta_{IR}\sigma_{IR}(t) = \int\limits_{\calL_{UV,t}}\Delta\sigma = \left(\frac\ii\hbar\right)^2 \int\limits_{\calL_{UV,t}}(T_{IR}+T_{UV})\,\sigma
 = \left(\frac\ii\hbar\right)^2 T_{IR}  \int\limits_{\calL_{UV,t}}\sigma,
 \]
 i.e.,
 \[
  \Delta_{IR}\sigma_{IR} (t) = \left(\frac\ii\hbar\right)^2 T_{IR}\,\sigma_{IR}(t),
 \]
 which shows that the relation is well-behaved under BV-pushforward (with the above assumptions).

 As for deformations of the $T_{UV}$\ndash Lagrangian submanifold, a direct computation using \eqref{e:dotIpf} simply yields
 \[
 \frac\dd{\dd t}\sigma_{IR}(t) = -\left(\frac\ii\hbar\right)^2 T_{IR} \int\limits_{\calL_{UV,t}}\dot\psi\,\sigma - \Delta_{IR}\int\limits_{\calL_{UV,t}} \dot\psi\,\sigma.
 \]
 In view of \eqref{e:Deltazero}, the $\Delta_{IR}$\ndash exact term 
%  change of $\sigma_1$
 is irrelevant in the BV formalism. As the other term is proportional to $T_{IR}$, it is also irrelevant, since we are going to consider only $T_{IR}$\ndash Lagrangian submanifolds for the further gauge fixing.
 
 If we now apply this to the equivariant BV case where $T$ is defined by \eqref{e:Tequiv} and $\calM_{IR}$ and $\calM_{UV}$ are based on AKSZ superfields, we see that the assumptions are satisfied and that the effective theory is precisely as described at the end of Section~\ref{s:equiBV}.  Namely, for the effective theory we still have 
 \[
 S^\text{eff}_{\Hat L_{v_a}} =  \int\limits_{\Sigma_d} {\mathbf P}_i\,L_{v_a} {\mathbf Q}^i
 \]
 where now the ${\mathbf P}$'s and ${\mathbf Q}$'s are  the superfields on $\calM_{IR}$. On the other hand, $S^\text{eff}_{\Hat \iota_{v_a}}$ will in general have a very complicated, nonlocal expression. 
 
 %{ \color{red}{define formally equivariant BV push-forward, list the conditions on the different structures, like 
 % invariance of ${\mathcal L}_{UV}$ etc and list its formal properties}}
 
 \section{BV formalism for DW}\label{s:DW}
 
 In this section we describe the AKSZ construction \cite{AKSZ:geometry_of_BV} for a
  four-dimensional field theory  to which we refer  as 
  the Donaldson-Witten (DW) theory since upon the appropriate gauge fixing it can be reduced to the standard cohomological 
   field theory described by Witten in \cite{w88}.  This DW BV theory has been discussed previously  in
    \cite{equiv-BV, Ca-1, CCFMRTZ, Ik} 
 
 Let $\mathfrak{g}$ be a finite dimensional quadratic  Lie algebra  with
  the Lie bracket $[~,~]$ and with the invariant metric $\langle~,~\rangle$.  
  Consider the graded vector space 
 \bea
  \mathfrak{g} [1] \oplus \mathfrak{g}[2]~,
 \eea
  which is equipped with the symplectic structure of degree $3$ 
  \bea
   \omega = \langle \delta c ,  \delta  \phi \rangle ~,
  \eea
  where $c$ is the coordinate of degree $1$ and $\phi$ is the coordinate of degree 2. We can introduce the Hamiltonian function of 
   degree $4$
   \bea
    \Theta = \frac{1}{2} \langle \phi, \phi \rangle + \frac{1}{2} \langle \phi, [c,c]\rangle~,
   \eea
  which generates the following odd vector field
 \bea
 && \targetq c = \phi + \frac{1}{2} [c, c]~,\\
 && \targetq \phi = [c, \phi]~,
 \eea
  such that $\targetq^2=0$. Following  the standard AKSZ construction \cite{AKSZ:geometry_of_BV}  we can define the space of maps
\begin{equation}
 \MDW= \{ ~ T[1]\Sigma_4~\longrightarrow~ \mathfrak{g} [1] \oplus \mathfrak{g}[2] ~ \}~,
  \end{equation}
 where $\Sigma_4$ is a $4$--manifold. Denoting by $\mathbf{C}$ and $\mathbf{\Phi}$ the superfields of degree $1$ and $2$, respectively, 
  the BV symplectic form is defined as
  \bea
   \omega_{DW} = \int\limits_{T[1]\Sigma_4} d^4x d^4\theta~ \delta \mathbf{C}^a \wedge \delta \mathbf{\Phi}^b \eta_{ab} \label{fullBV-sympl}
  \eea
  and the AKSZ action as
  \begin{equation}
   \SDW = \int\limits_{T[1]\Sigma_4} d^4x d^4\theta~ \left ( \langle\mathbf{\Phi},  d \mathbf{C} \rangle + \frac{1}{2} \langle \mathbf{\Phi}, \mathbf{\Phi} \rangle 
   + \frac{1}{2} \langle \mathbf{\Phi}, [\mathbf{C},\mathbf{C}]\rangle \right )~.\label{fullBV-action}
  \end{equation}
  %where we split two forms into self-dual and anti-self-dual by introducing the metric on $\Sigma_4$.  
   The BV transformations are 
   \bea
  &&  \sfdelta \mathbf{C} = d\mathbf{C} + \mathbf{\Phi} + \frac{1}{2} [\mathbf{C}, \mathbf{C}]~,\\
   && \sfdelta \mathbf{\Phi} = d \mathbf{\Phi} + [\mathbf{C}, \mathbf{\Phi}]~.
   \eea
    
  We adopt the following convention for the expansion of superfields 
  in terms of differntial forms
  \bea
  &&\mathbf{C} = c_0 + c_1 + c_2 + c_3 + c_4   ~,\\
  &&\mathbf{\Phi} = \phi_0 + \phi_1 + \phi_2 + \phi_3 + \phi_4 ~.
  \eea
The BV symplectic structure can then be written as
  \begin{equation}
    \omega_{DW} = \int\limits_{\Sigma_4} \left ( \delta c_0 \wedge \delta \phi_4 + \delta c_1 \wedge \delta \phi_3 + \delta c_2 \wedge \delta \phi_2
    + \delta c_3 \wedge \delta \phi_1 + \delta c_4 \wedge \delta \phi_0 \right )~,
  \end{equation}
and the action (\ref{fullBV-action}) reads
   \begin{eqnarray}\label{AKSZ_components}
  \SDW &=& \int \langle \phi_0, d_{c_1} c_3 \rangle + \langle \phi_1, d_{c_1} c_2 \rangle + \langle \phi_2, F(c_1)\rangle + \langle \phi_3 , d_{c_1} c_0 \rangle\cr
  & &  +   \langle \phi_0, \phi_4 \rangle + \langle \phi_1, \phi_3 \rangle + \frac{1}{2} \langle \phi_2, \phi_2 \rangle + \langle \phi_0, [c_0, c_4]\rangle \cr
  & & + \frac{1}{2} \langle \phi_0, [c_2, c_2]\rangle+ \langle \phi_1, [c_0, c_3]\rangle + \langle \phi_2, [c_0, c_2] \rangle + \frac{1}{2} \langle \phi_4, [c_0, c_0]\rangle
   \end{eqnarray}
   where $d_{c_1} = d + [c_1,~]$ and $F(c_1) = dc_1 + \frac{1}{2} [c_1, c_1]$. Restricting to the physical ({\it i.e. degree $0$}) fields
   \begin{equation}
    \SDW|_{cl}= \int \langle \phi_2, F(c_1) \rangle + \frac{1}{2} \langle \phi_2, \phi_2 \rangle~,
   \end{equation}
    where upon the integration of $\phi_2$ we obtain the standard topological term {
    %7 
    $-\frac{1}{2}\int F(c_1)^2$ } on which the original treatment of the  DW theory is based.

The 4D AKSZ action (\ref{fullBV-action}) can also be encoded in the following space
   \bea
   T[1]\Sigma_4 \oplus \mathbb{R}[-1] ~\longrightarrow~ \mathfrak{g} [1]  ~,
  \eea
where we denote $\theta$'s as odd coordinates on $T[1]\Sigma_4$ and $\xi$ as an odd coordinate on $\mathbb{R}[-1]$. 
  The new superfield is written in terms of old as follows
  \bea\label{A_superfield}
   \mathbf{A} = \xi \mathbf{\Phi} +  \mathbf{C} 
  \eea
 The symplectic structure is 
   \bea
   \omega_{DW} = \int\limits_{T[1]\Sigma_4\oplus \mathbb{R}[-1] } d^4x d^4\theta d\xi ~ \delta \mathbf{A}^a \wedge \delta \mathbf{A}^b \eta_{ab} \label{5D-fullBV-sympl}
  \eea
and 
the AKSZ action becomes
 \begin{equation}\label{DW_CS_form}
   \SDW = \int\limits_{T[1]\Sigma_4 \oplus \mathbb{R}[-1] } d^4x d^4\theta d\xi ~ \left ( \langle \mathbf{A}, (d + \frac{\partial}{\partial \xi}) \mathbf{A}\rangle + \mathbf{A}\mathbf{A}\mathbf{A}\right )
   \end{equation}
   and it reduces to (\ref{fullBV-action}). It is a Chern-Simons action with source the dg-manifold $(T[1]\Sigma_4\times{\mathbb R}[-1],D+\partial/\partial\xi)$ and degree $-3$ integral $\int=\int d^4x d^4\theta d\xi$; the dga of global functions is then $(\Omega(\Sigma_4)[\xi],D+\frac{\partial}{\partial\xi})$. The underlying complex can be described as   
   
   \begin{equation}\label{DW_dGA_CSform}
\begin{tikzcd}[sep=small]
 & \Omega^0 (\Sigma_4) \arrow[r] & \Omega^1 (\Sigma_4) \arrow[r] & \Omega^{2}(\Sigma_4) \arrow[r]    & \Omega^3(\Sigma_4) \arrow[r]    & \Omega^4(\Sigma_4)   \\
   \Omega^0  (\Sigma_4) \arrow[ur] \arrow[r]   &   \Omega^1 (\Sigma_4)  \arrow[ur] \arrow[r]   & \Omega^{2} (\Sigma_4)  \arrow[r] \arrow[ur]   &  \Omega^3 (\Sigma_4)  \arrow[r] \arrow[ur]   &  \Omega^4 (\Sigma_4)   \arrow[ur]  &
     \end{tikzcd}
        \end{equation}
where the horizontal arrows denote de Rham differential $d$ and the diagonal arrows denote $\partial/\partial\xi$. 
The field components are defined by expanding the superfield $\mathbf{A}$ along  the following basis of local coordinates (as a $\Omega^0(\Sigma_4)$-module) 
   \bea
   && \xi~,~~~{\rm degree~-1} \nonumber\\
   && 1~,~~~\theta^i\xi~,~~~~{\rm degree~0} \nonumber\\
   && \theta^i~,~~~\theta^i\theta^j\xi~,~~~~{\rm degree~1}  \nonumber \\
   && \theta^i\theta^j~,~~~\theta^i\theta^j\theta^k\xi~,~~~~{\rm degree~2} \nonumber\\
   && \theta^i\theta^j \theta^k~,~~~\theta^1\theta^2\theta^3\theta^4\xi~,~~~~{\rm degree~3}\nonumber\\
   &&\theta^1\theta^2\theta^3\theta^4~,~~~~{\rm degree~4}~~~~~~.\nonumber
   \eea

  \section{BV formalism for YM}\label{s:YM}
  Let us introduce a metric on $\Sigma_4$ and let us decompose the two form components in self-dual and anti-self-dual forms as
  \begin{equation}
   \phi_2 = \phi_2^+ + \phi_2^-~,~~~~~c_2 = c_2^{+} + c_2^{-}~.
  \end{equation}
If we denote with 
\begin{equation}\label{IR_fields}
\MIR=\{c_2^-=c_3=c_4=\phi_0=\phi_1=\phi_2^-=0\}
\end{equation} 
and
\begin{equation}\label{UV_fields}
\MUV=\{c_0=c_1=c_2^+=\phi_2^+=\phi_3=\phi_4=0\} 
\end{equation}
we see that $\MDW$ is decomposed in the direct product of symplectic manifolds
\begin{equation}\label{odd_sympl_product}
  \MDW =  \MIR \times  \MUV\;,
\end{equation}
where the symplectic forms read as  
   \begin{equation}
    \omega_{IR} = \int\limits_{\Sigma_4} \left ( \delta c_0 \wedge \delta \phi_4 + \delta c_1 \wedge \delta \phi_3 + \delta c_2^+ \wedge \delta \phi_2^+   \right )~,
  \end{equation}
and 
     \begin{equation}
    \omega_{UV} = \int\limits_{\Sigma_4} \left (  
    \delta c_2^- \wedge \delta \phi_2^- + \delta c_3 \wedge \delta \phi_1 + \delta c_4 \wedge \delta \phi_0 \right )~.
  \end{equation}
As in Sections \ref{s:BV-overview} and \ref{s:equiv-BV-overview}
we are using the subscripts $IR$ and $UV$ that stand for {\it infrared} and {\it ultraviolet} respectively. 
  
Restricting $\SDW$ to $\MIR$ 
(i.e., setting to zero $\phi_2^-$, $c_2^-$, $\phi_1$, $c_3$, $\phi_0$ and $c_4$ in (\ref{AKSZ_components})) we get
\begin{eqnarray}\label{YM_action}
 \SIR &\equiv& \SDW|_{{\mathcal M}_{IR}}=\int \langle \phi_2^+, F(c_1) \rangle + \langle \phi_3, d_{c_1} c_0 \rangle 
+ \frac{1}{2} \langle \phi_2^+, \phi_2^+ \rangle \cr
& & \phantom{S_{AKSZ}|_{{\mathcal M}_{YM}}=\int} + \langle \phi_2^+, [c_0, c_2^+]\rangle 
   + \frac{1}{2} \langle \phi_4, [c_0, c_0]\rangle\;.
\end{eqnarray}
  
It is crucial to observe that we obtain the same formula if we restrict $\SDW$ to $\MIR\times{\mathcal L}_{UV}$, where ${\mathcal L}_{UV}$ is the lagrangian submanifold of $\MUV$ defined by $\phi_0=\phi_1=\phi_2^-=0$ ({\it i.e.} keeping $c_2^-,c_3$ and $c_4$ different from zero). It is easy to see that this implies that 
\begin{equation}\label{CME_YM}
 \{\SIR,\SIR\} = 0\;.
\end{equation}
By restricting (\ref{YM_action}) further to the physical ({\it i.e.} degree $0$) fields we get 
$$
\SIR|_{cl} = \int\limits_{\Sigma_4} \langle \phi_2^+, F(c_1) \rangle+ \frac{1}{2} \langle \phi_2^+, \phi_2^+ \rangle \equiv S_{YM}^{fo}
$$
that is the first order formulation of the Yang-Mills action. Upon the integration of $\phi_2^+$ we obtain the standard Yang-Mills action (up to topological term).   
This proves that $\SIR$ is a BV extension of the Yang-Mills action; we are going to prove that it coincides with Costello's solution
in Chapter 6.2.1 of \cite{Costello}. A similar result can be found also in \cite{CCFMRTZ}.

\subsection{Costello's formulation of Yang-Mills}\label{CostelloYM}
Let us consider now the Chern-Simons formulation of $\SDW$ written in (\ref{DW_CS_form}). The restriction to $\MIR$ imposes constraints to the superfield (\ref{A_superfield}) so that $S_{IR}$ is not anymore in the AKSZ form; let us see how it can still be put in this form. 

Let us consider ${\mathcal A}_0=\{\omega+(\nu_2^++\nu_3+\nu_4)\xi\}$ that is a dg-subalgebra of $(\Omega(\Sigma_4)[\xi],d+\frac{\partial}{\partial\xi})$ together with its dg-ideal ${\mathcal I}_0=\{\omega_2^-+\omega_3+\omega_4\}$ (recall that $\nu_2^+\wedge\omega_2^-=0)$. The quotient dga
\begin{equation}\label{costello_dGA}
({\mathcal A} = {\mathcal A}_0/{\mathcal I}_0, d+\frac{\partial}{\partial\xi} )
\end{equation}
encodes the following complex 
   \begin{equation}\label{YM_source}
\begin{tikzcd}
  \Omega^0 (\Sigma_4) \arrow[r] & \Omega^1 (\Sigma_4) \arrow[r] & \Omega^{2+}(\Sigma_4)      &     \\
    & \Omega^{2+} (\Sigma_4)  \arrow[r] \arrow[ur]   &  \Omega^3 (\Sigma_4)  \arrow[r] &  \Omega^4 (\Sigma_4)    
     \end{tikzcd}
        \end{equation}
where the horizontal map is the de Rham differential (composed with the projection to self dual forms in degree $1$ and to $0$ in degree $2$ of the upper line) and the diagonal one is the identity. Analogously, the algebra structure is the wedge product composed with the same projections when it involves forms of the upper line. By inspection we check that in (\ref{YM_action}) there is no difference if we assume that forms are multiplied according to this exotic ring structure instead of just with the wedge product. We then showed that also (\ref{YM_action}) is an AKSZ solution of the master equation {%8 
(\ref{CME})} 
with target $\g[1]$ and with source the (spec of the) 
dg-algebra $({\mathcal A},d+\frac{\partial}{\partial\xi})$ endowed with the degree $-3$ integral 
$\int:{\mathcal A}\rightarrow \R$ defined as
$$
\int(\omega+\nu\xi) = \int\limits_{\Sigma_4} \nu\;.
$$
A local basis of $\mathcal A$ (as a $\Omega^0(\Sigma_4)$-module) is 
   \bea\label{basis_costello_algebra}
   && 1~,~~~~{\rm degree~0} \\
   && \theta^i~,~~~\pi^+(\theta^i\theta^j)\xi~,~~~~{\rm degree~1}  \nonumber \\
   && \pi^+(\theta^i\theta^j)~,~~~\theta^i\theta^j\theta^k\xi~,~~~~{\rm degree~2} \nonumber\\
   && \theta^1\theta^2\theta^3\theta^4\xi~,~~~~{\rm degree~3}\nonumber
   \eea
   where $\pi^+$ is the projection on self-dual two forms. 
We can summarize by saying that
\begin{equation}\label{YM_CS_form}
   \SIR = \int ~   \langle \mathbf{A}, (d + \frac{\partial}{\partial \xi}) \mathbf{A}\rangle + \mathbf{A}\mathbf{A}\mathbf{A} ~,
   \end{equation}
where now $\mathbf{A}$ is expanded along the local basis (\ref{basis_costello_algebra}) and the product is taken in $\mathcal A$.    
This CS solution of the master equation for Yang-Mills was first introduced in Lemma 2.1.1 of \cite{Costello}. In the following sections we will refer to $(\A,d+\frac{\partial}{\partial\xi})$ as the Costello dg-algebra.

The extension of the AKSZ construction to the case when the source dg algebra is not locally freely generated (and so is not the algebra of functions on a dg manifold) has been discussed in \cite{BMZ}.

\subsection{BV-pushforward} 
The decomposition (\ref{odd_sympl_product}) defines also the (trivial) fibration ${\mathcal M}_{DW}\rightarrow{\mathcal M}_{IR}$ where we interpret 
\begin{equation}\label{IR_fields}IR\equiv\{\Icz,\Ico,\Ictwop,\Iphitwop,\Iphithree,\Iphif\}\in {\mathcal M}_{IR}
\end{equation}
 
as {\it infrared fields} and 
\begin{equation}\label{UV_fields}UV\equiv\{\Uctwom,\Ucthree,\Ucf,\Uphiz,\Uphio,\Uphitwom\}\in{\mathcal M}_{UV}
\end{equation}
 
as {\it ultraviolet fields}. We are then in a position to define the BV pushforward of the topological DW theory to the sector of the physical YM fields.

Let us decompose the full BV action (\ref{AKSZ_components}) as
$$
\SDW = \SIR + \SUV +S_{mixed}
$$
where 
\begin{equation}\label{UV_action}
  \SUV = \int \langle \Uphiz, d \Ucthree \rangle + \langle \Uphio, d \Uctwom \rangle 
 + \frac{1}{2} \langle \Uphitwom, \Uphitwom \rangle + \frac{1}{2} \langle \Uphiz, [\Uctwom, \Uctwom]\rangle 
   \end{equation}
and
\begin{eqnarray}\label{mixed_UV_IR}
  S_{mixed} &=& \int \langle \Uphiz, [\Ico, \Ucthree] \rangle + \langle \Uphio, d_{\Ico} \Ictwop \rangle +\langle\Uphio,[\Ico,\Uctwom]\rangle +\langle \Uphitwom, F(\Ico)\rangle \cr
  & &  + \langle \Uphiz, \Iphif \rangle + \langle \Uphio, \Iphithree \rangle +  \langle \Uphiz, [\Icz, \Ucf]\rangle + \langle \Uphio, [\Icz, \Ucthree]\rangle \cr
  & & + \langle \Uphitwom, [\Icz, \Uctwom] \rangle+ \frac{1}{2} \langle \Uphiz, [\Ictwop, \Ictwop]\rangle \rangle\;.
   \end{eqnarray}

In order to perform the vertical integration it is necessary to separate the UV fields in cohomology and fluctuations, by using the Hodge %{9 s}  
decomposition of forms. Let us define 
\begin{equation}\label{UV_zeromodes}
{\color{blue} c}={\color{blue}\tilde{c}}+{\color{red} h}\,\;\;\;{\color{blue}\phi}={\color{blue}\tilde{\phi}}+{\color{red}\sigma}
\end{equation}
where ${\color{red}h}\equiv\{{\color{red}h_2^-},{\color{red}h_3},{\color{red}h_4}\}=P({\color{blue}c}),{\color{red}\sigma}\equiv\{{\color{red}\sigma_0},{\color{red}\sigma_1},{\color{red}\sigma_2^-}\}=P({\color{blue}\phi})$ are harmonic forms ($P$ denotes the projection to harmonic forms) and ${\color{blue}\tilde{c}},{\color{blue}\tilde{\phi}}$ denote the fluctuations. We call ${\color{red}\sigma},{\color{red}h}$ the {\it zero modes}; they are colored in red because, like the IR fields, they are not integrated out and must be treated as external fields. The zero modes symplectic form reads
$$
\int (\delta {\color{red}\sigma_0}\wedge\delta {\color{red}h_4 }+\delta{\color{red}\sigma_1}\wedge\delta {\color{red}h_3} + \delta{\color{red}\sigma_2^-} \wedge\delta {\color{red}h_2^-})\;. 
$$ 
The gauge fixing is now imposed on the UV-fluctuations by choosing a Lagrangian submanifold ${\mathcal L}_{UV}$ defined by
\begin{equation}
 \label{UV_gauge_fixing}
 {\color{blue}\tilde{c}_4}={\color{blue}\tilde{\phi}_2^-}=0\;,\;\;\; {\color{blue}\tilde{c}_3},{\color{blue}\tilde{\phi}_1}\in{\rm Im} d^\dagger\;,
\end{equation}
where $d^\dagger$ denotes the codifferential. 
The effective action is now defined by
$$
e^{\frac{i}{\hbar}S_{\text{eff}}}= \int\limits_{{\mathcal L}_{UV}} e^{\frac{i}{\hbar}S_{DW}}\;.
$$

Let us analyze the abelian case first. The explicit computation is postponed to the next section where the more general equivariant case will be analyzed. We anticipate here that in the non equivariant abelian case the only corrections to (\ref{YM_action}) come from the UV zero modes contributions, {\it i.e.}
\begin{equation}\label{effective_action_nonequiv}
S_{\text{eff}} = S_{IR} + \int \langle \sigma_0,\phi_4\rangle + \langle \sigma_1,\phi_3\rangle + \frac{1}{2} \langle \sigma_2^-,\sigma_2^-\rangle
\end{equation}
where the last terms come from the evaluation of (\ref{UV_action}) and (\ref{mixed_UV_IR}) on zero modes. Since all remaning fields are red, we suppressed the coloring. The abelian YM-BV differential is corrected to
\begin{equation}\label{abelian_non_equiv_differential_IR}
\sfdelta c_0 = \sigma_0,\;\;\;
\sfdelta c_1 = d c_0 +\sigma_1,\;\;\;
\sfdelta c_2^+ = \pi^+d c_1 + \phi_2^+ 
\end{equation}
$$
\sfdelta \phi_2^+ = 0,\;\;\; 
\sfdelta \phi_3 = d\phi_2^+ ,\;\;\;
\sfdelta \phi_4 = d \phi_3 
$$
The UV zero modes are not integrated out so far and so they survive as dynamical variables. In particular we have
$$
\sfdelta \sigma_0 = \sfdelta \sigma_1 = \sfdelta \sigma_2^- =0
$$
\begin{equation}\label{non_equivariant_differential_ZM}
\sfdelta h_2^- =\sigma_2^- \,,\;\; \;\;
\sfdelta h_3 = P\phi_3\,,\;\;\;\; \sfdelta h_4 = P \phi_4\;.
\end{equation}
By construction $\sfdelta$ satisfies
$$
\sfdelta^2 = 0\;.
$$
\begin{rmk}\label{enlarged_costello}
It is easy to check that the effective action (\ref{effective_action_nonequiv}) is the CS action (\ref{YM_CS_form}) with the replacement ${\mathbf A}\rightarrow{\mathbf A}+h+\xi \sigma$, where $ h = h_2^-+h_3+h_4$ and ${\mathbf \sigma}=\sigma_0+\sigma_1+\sigma_2^-$ are the superfields encoding the UV zero modes. 
% In other terms (\ref{effective_action_nonequiv}) is the AKSZ action with target $\g[1]$ and with source the (spec of) the Costello algebra ${\mathcal A}$ times the space of UV zero modes.
The homological vector field $d+\frac{\partial}{\partial\xi}$ is encoded in the following complex that is an enlarged version of (\ref{YM_source})
\begin{equation}\label{d-1-diagram}
\begin{tikzcd}
 & \Omega^0  \arrow[r, "d"] & \Omega^1 \arrow[r, "\pi^+ d \oplus 0"] & \Omega^{2+} \oplus H^{2-}     & H^3    & H^4   \\
   H^0 \arrow[ur, "i"]   &   H^1 \arrow[ur, "i"]    & \Omega^{2+}\oplus H^{2-} \arrow[r, "d\oplus 0"] \arrow[ur, "id"]   &  \Omega^3 \arrow[r, "d"] \arrow[ur, "P"]   &  \Omega^4  \arrow[ur, "P"]  &
     \end{tikzcd}
        \end{equation}
    where $i$ denotes the embedding of de Rham cohomology into differential forms as harmonic forms. %The ring structure is 
%     ${\mathcal A}\otimes_{\R[\xi]}{\mathcal Z}_{UV}$ where ${\mathcal Z}_{UV}=H^{2-}\oplus H^3\oplus H^4\oplus(H^0\oplus H^1\oplus H^{2-})[\xi]$ with ring structure 
%    given by the wedge product composed with the projections to cohomology and antiself dual forms.
 \end{rmk}
 \begin{rmk} The calculation of $S_{\text{eff}}$ for non-abelian theory involves all orders in the perturbation theory and will be discussed in a future paper. %{\color{red} Can we say something about $(S_{eff})|_{cl}$ ?} %However on the general ground we will get the deformations of $d_{+1}$ differential which still will square to zero. However the correction will be non-local and they will not be a derivations with respect to the ring structure. 
 \end{rmk}

\section{equivariant BV pushforward map}\label{s:equiv-DW-YM}

In the previous section we have constructed 4d DW theory as the standard AKSZ theory with the source being $T[1]\Sigma_4$ and the target 
  $\mathfrak{g} [1] \oplus \mathfrak{g}[2]$.  Let us consider now a vector field $v\in\Gamma(T\Sigma_4)$ (possibly a complete one whose flux defines a group action) and let us consider the equivariant extension of the ASKZ action as describd in Section \ref{s:equiv-BV-overview}.
  
 Borrowing the notation from the previous section, the equivariant AKSZ action is defined on the same space of fields $\MDW$ as
  \begin{eqnarray}
S_{eq-DW}  &=& \SDW+S_{\Hat\iota_v}\cr
&=&\int\limits_{T[1]\Sigma_4} d^4x d^4\theta~ \left ( \langle\mathbf{\Phi},  d_v \mathbf{C} \rangle + \frac{1}{2} \langle \mathbf{\Phi}, \mathbf{\Phi} \rangle 
   + \frac{1}{2} \langle \mathbf{\Phi}, [\mathbf{C},\mathbf{C}]\rangle \right )~,    \label{fullequivBV-action}
  \end{eqnarray}
where $d_v= d +\iota_v$ is the (Cartan) equivariant differential  and 
\begin{equation}\label{hamiltonian_contraction}
S_{\Hat \iota_v} = \int\limits_{T[1]\Sigma_4} d^4x d^4\theta~   \langle\mathbf{\Phi},  \iota_v \mathbf{C} \rangle 
\end{equation}
is the hamiltonian defining the lift $\Hat\iota_v$ of the contraction vector field $\iota_v$ to $\MDW$. 
 For the clarity of forthcoming  formulas here we suppress the formal variable $u$ and assume that we have just one $v$, in the final result is easy to 
  restore $u^a$ and $v_a$. 
This action satisfies the equivariant master equation 
\begin{equation}\label{equiv_CME}
    \frac{1}{2}\{ S_{eq-DW}, S_{eq-DW} \} =- S_{\Hat L_v} ~,
\end{equation}
where
\begin{equation}\label{hamiltonian_lie_derivative}
S_{\Hat L_v}
= \int\limits_{T[1]\Sigma_4} d^4x d^4\theta~   \langle\mathbf{\Phi},  L_v \mathbf{C} \rangle
\end{equation}
is the hamiltonian defining the lift $\Hat{L}_v$ of ${L}_v$ to $\MDW$.  In components $S_{eq-DW}$ is given by
 \begin{equation}\label{fullBVaction}
    S_{eq-DW} = \SDW + \int  \langle \phi_1,  \iota_v c_4  \rangle +  \langle \phi_2,  \iota_v c_3 \rangle
      + \langle \phi_3,  \iota_v c_2 \rangle   +  \langle \phi_4,  \iota_v c_1  \rangle
    \end{equation}
     where the component expression of $S_{DW}$ is given in  (\ref{AKSZ_components}).

Let us choose now a metric on $\Sigma_4$ invariant under $L_v$ and consider the splitting $\MDW = {\mathcal M}_{IR} \times {\mathcal M}_{UV}$ defined in (\ref{odd_sympl_product}). 
The restriction  $S_{eq-DW}$ to $\MIR$ does not satisfy the equivariant master equation (\ref{equiv_CME}) (since the lift $\Hat\iota_v$ to $\MDW$ of the contraction operator $\iota_v\in\Der(\Omega(\Sigma_4)[\xi])$ does not restrict to $\MIR$). In other terms, the invariance of the metric guarantees that $\MIR$ inherits from $\MDW$ the $\Hat{L}_v$ action but not the Cartan complex for equivariant cohomology. 

\begin{rmk}
We saw in Section \ref{CostelloYM} that the infrared action $S_{IR}$, which is a BV extension of Yang-Mills, is an AKSZ action by itself with source the dg algebra $\mathcal A$ defined in (\ref{costello_dGA}). It is natural then to look for the definition of the Cartan complex for equivariant cohomology on $\mathcal A$ and then lifting it to $\MIR$. It is easy to check that while the Lie derivative ${L}_v\in\Der(\Omega(\Sigma)[\xi])$ descends to $\Der({\mathcal A})$, the same is not true for $\iota_v\in\Der(\Omega(\Sigma)[\xi])$. So there is not a natural notion for the contraction operator.
\end{rmk}

We are going instead to compute the BV push-forward map for $S_{eq-DW}$
       to the space $\MIR$; by construction the effective action will satisfy the equivariant master equation. 

The general non-abelian theory would require the full perturbation theory. At the same time in the abelian case the calculation can be carried out exactly still being non-trivial. Thus from now on we switch to the abelian equivariant DW theory. 
     
   Borrowing the notations from the previous section we can split the abelian $S_{eq-DW}$ as follows
     \begin{equation}
       S_{eq-DW} = S_{IR} + S_{UV} + S_{mixed}
     \end{equation}
    with $\{ \Icz,\Ico,\Ictwop,\Iphitwop,\Iphithree,\Iphif \} \in {\mathcal M}_{IR}$
    \begin{equation}
     S_{IR} = \int \left ( \Iphitwop d \Ico +  \Iphithree d \Icz + \frac{1}{2} (\Iphitwop)^2 + \Iphithree \iota_v \Ictwop + \Iphif  \iota_v \Ico \right )
    \end{equation}
    and $\{\Uctwom,\Ucthree,\Ucf,\Uphiz,\Uphio,\Uphitwom\}\in{\mathcal M}_{UV}$
    \begin{equation}
     S_{UV} = \int \left ( \Uphiz d \Ucthree + \Uphio d  \Uctwom  + \frac{1}{2} (\Uphitwom)^2 +   \Uphio \iota_v \Ucf + \Uphitwom \iota_v \Ucthree  \right )  
    \end{equation}
     and 
     \begin{equation}\label{mixed}
  S_{mixed} = \int \left ( \Uphio d \Ictwop +   \Uphitwom d \Ico       + \Uphiz \Iphif   + \Uphio \Iphithree + \Iphithree \iota_v \Uctwom + \Iphitwop \iota_v \Ucthree \right ) \\
 \end{equation}    
   We stress that we use the ${L}_v$-invariant metric to define all fields. 

We decompose the UV fields as in (\ref{UV_zeromodes}) by using the Hodge decomposition with respect to the invariant metric. We then define the lagrangian submanifold ${\mathcal L}_{UV}$ as follows
     \begin{equation}\label{UV-Lag-def}
      {\color{blue}\tilde{c}_4}={\color{blue}\tilde{\phi}_2^-}=0\;,\;\;\; {\color{blue}\tilde{c}_3},{\color{blue}\tilde{\phi}_1}\in{\rm Im}\, d^\dagger\;,
     \end{equation}
    and by construction ${\mathcal L}_{UV}$ is invariant under the action of ${L}_v$.  Before imposing the conditions ${\color{blue}\tilde{c}_3},{\color{blue}\tilde{\phi}_1}\in{\rm Im}\,  d^\dagger$ explicitly let us rewrite a bit $S_{IR}$, $S_{UV}$ and $S_{mixed}$ assuming just that ${\color{blue}\tilde{c}_4}={\color{blue}\tilde{\phi}_2^-}=0$.  Assuming the decomposition (\ref{UV_zeromodes}) $S_{UV}$ is written as 
           \begin{equation}
     S_{UV} = \int \left ( {\color{blue}\tilde{\phi}_0} d {\color{blue}\tilde{c}_3} + {\color{blue}\tilde{\phi}_1} d 
     {\color{blue}\tilde{c}_2^-}  + \frac{1}{2} ({\color{red}\sigma_2^-})^2 +{\color{blue}\tilde{\phi}_1} \iota_v {\color{red} h_4 }+ 
     {\color{red} \sigma_2^- }\iota_v {\color{blue} \tilde{c}_3} 
     + {\color{red} \sigma_1} \iota_v {\color{red} h_4} + {\color{red} \sigma_2^-} \iota_v {\color{red} h_3} \right )  
    \end{equation}
      and $S_{mixed}$ as follows
      \begin{eqnarray}
  &S_{mixed} = & \int \left ({\color{blue} \tilde{\phi}_1} d{\color{red}c_2^+}  + {\color{blue} \tilde{\phi}_0} {\color{red} \phi_4} + {\color{red}\sigma_0}{\color{red} \phi_4}
    + {\color{blue} \tilde{\phi}_1} {\color{red} \phi_3} + {\color{red} \sigma_1}{\color{red} \phi_3}  +
     {\color{red} \phi_3} \iota_v {\color{blue} \tilde{c}_2^-} + {\color{red} \phi_3} \iota_v {\color{red}\sigma_2^-}  \right . \nonumber \\
   && \left .  + {\color{red} \phi_2^+} \iota_v {\color{blue} \tilde{c}_3}  + {\color{red} \phi_2^+} \iota_v {\color{red} h_3}       \right ) ~.
 \end{eqnarray}
  We are going to integrate out only the blue fields  $\{{\color{blue}\tilde{\phi}_0}, {\color{blue}\tilde{c}_3}, {\color{blue}\tilde{\phi}_1}, 
     {\color{blue}\tilde{c}_2^-}\}$ and the red fields (IR fields and UV zero modes) are treated as external.  Therefore it is useful to identify 
      $S_0$ as containing only quadratic terms in the blue fields 
     \begin{equation}\label{modified-SUV}
     S_0 = \int \left ( {\color{blue}\tilde{\phi}_0} d {\color{blue} \tilde{c}_3} + {\color{blue}\tilde{\phi}_1} d  {\color{blue}\tilde{c}_2^-} \right ) =  \int \left ( \frac{1}{2} {\color{blue}\tilde{\phi}_0} d {\color{blue} \tilde{c}_3} + \frac{1}{2} {\color{blue}\tilde{c}_3} d 
     {\color{blue} \tilde{\phi}_0} + \frac{1}{2}  {\color{blue}\tilde{\phi}_1} d  {\color{blue} \tilde{c}_2^-}
      + \frac{1}{2} {\color{blue}\tilde{c}_2^-} (\pi^- d ) {\color{blue} \tilde{\phi}_1}   \right ) ~,
    \end{equation}
     where $\pi^\pm$ are the projectors on $\Omega^{2\pm}$, $S_1$ as containing the linear terms in the blue fields
    \begin{equation}\label{mixed-first}
       S_1  = \int \left ( {\color{blue} \tilde{\phi}_0} {\color{red} \phi_4} - {\color{blue}\tilde{c}_3} ( \iota_v {\color{red}\phi_2^+}  + \iota_v {\color{red}\sigma_2^-}) + {\color{blue}\tilde{\phi}_1} ({\color{red}\phi_3} + d {\color{red} c_2^+} + \iota_v {\color{red}h_4})  + {\color{blue}\tilde{c}_2^-} \iota_v 
       {\color{red}\phi_3} \right )
\end{equation}
 and new  $\tilde{S}_{IR}$ containing only red fields (including UV zero modes)
\begin{eqnarray}\label{IR-term-mixed}
  &  \tilde{S}_{IR} &= \int \left ( {\color{red}\phi_2^+} d {\color{red} c_1} + {\color{red} \phi_3} d {\color{red} c_0} + \frac{1}{2} ({\color{red}\phi_2^+})^2 + 
  {\color{red}\phi_3} \iota_v {\color{red} c_2^+} + {\color{red} \phi_4} \iota_v {\color{red} c_1} 
     +   \frac{1}{2} ({\color{red}\sigma_2^-})^2  \right . \nonumber \\
  && \left .  + {\color{red} \sigma_1} \iota_v {\color{red} h_4} + {\color{red} \sigma_2^-} \iota_v {\color{red} h_3}
   + {\color{red}\sigma_0}{\color{red} \phi_4} + 
      {\color{red} \sigma_1}{\color{red} \phi_3} + {\color{red} \phi_3} \iota_v {\color{red} \sigma_2^-} + {\color{red} \phi_2^+} \iota_v {\color{red} h_3}
     \right )\;,
    \end{eqnarray}
so that $S_{eq-DW}= S_0+S_1+ \tilde{S}_{IR}$. It is convenient to write the quadratic form appearing in $S_0$ as 
     \bea   \label{quadratic_form}
   \frac{1}{2}  \left (
     \begin{array}{cccc}
     {\color{blue} \tilde{\phi}_0} & {\color{blue}\tilde{c}_3} & {\color{blue} \tilde{\phi}_1} & {\color{blue} \tilde{c}_2^-  }
     \end{array}
     \right )  
      \left( 
\begin{array}{cccc}
    0 &  d & 0  & 0 \\
  d & 0 & 0  & 0 \\
   0 & 0 & 0 &  d \\
  0 & 0 & (\pi^- d)  & 0
\end{array}
\right)
 \left (
     \begin{array}{c}
   {\color{blue}  \tilde{\phi}_0} \\
     {\color{blue}  \tilde{c}_3 }\\
      {\color{blue} \tilde{\phi}_1} \\
        {\color{blue} \tilde{c}_2^-}  
     \end{array}
     \right )  
\eea
 where the wedge product is assumed. 
Analogously the integrand in $S_1$ is written as 
  \bea
  \left (
     \begin{array}{cccc}
   {\color{blue}  \tilde{\phi}_0} & {\color{blue} \tilde{c}_3} & {\color{blue} \tilde{\phi}_1} &  {\color{blue} \tilde{c}_2^- } 
     \end{array}
     \right )  
    (1-P) \left (
     \begin{array}{c}
     {\color{red} \phi_4} \\
    - \iota_v {\color{red} \phi_2^+} - \iota_v{\color{red} \sigma_2^-} \\
     {\color{red} \phi_3} + d {\color{red} c_2^+} + \iota_v {\color{red} h_4}\\
     \pi^- \iota_v {\color{red}\phi_3}
     \end{array} \right ) 
  \eea
   where the wedge product is assumed and $(1-P)$ is the project outside of harmonic forms (cohomology). The claim is that the quadratic form in (\ref{quadratic_form}) is non degenerate 
   once  the gauge fixing conditions (\ref{UV-Lag-def}) are taken into account so that the UV integration is well defined. This will be discussed in the next subsection.
  
  \subsection{Integration of UV sector}  

 The canonical way to restrict fields to ${\mathcal L}_{UV}$ in (\ref{UV-Lag-def}) is by adding a trivial sector of auxiliary fields. Indeed, let us add $({\color{blue}\varphi},\color{blue}\varphi^{\vee},\color{blue} \eta,\color{blue} \eta^{\vee})$ with $\deg {\color{blue}\varphi} = -2$, $\deg {\color{blue}\varphi^{\vee}} =1$, 
    $\deg {\color{blue} \eta} = -1$ and $\deg {\color{blue} \eta^{\vee}} =0$, with ${\color{blue}\varphi}$ being (even) zero form
     (${\color{blue}\varphi^{\vee}}$ is top form) and ${\color{blue}\eta}$ being (odd) zero form (${\color{blue}\eta^{\vee}}$ top form).  The BV symplectic form in UV sector is now
  \begin{equation}
    \tilde{\omega}_{UV} = \int \left (\delta {\color{blue} \tilde{\phi}_2^-}  \wedge \delta {\color{blue}\tilde{c}_2^-} + 
    \delta {\color{blue} \tilde{\phi}_1} \wedge \delta {\color{blue} \tilde{c}_3} + \delta {\color{blue} \tilde{\phi}_0}
     \wedge \delta {\color{blue} \tilde{c}_4}  + \delta {\color{blue} \varphi} \wedge \delta {\color{blue} \varphi^{\vee}}
    + \delta {\color{blue}\eta} \wedge \delta {\color{blue} \eta^{\vee}} 
     \right)
   \end{equation}
    and the BV action is extended to  
    \begin{equation}
     S_{eq-DW} + \int \left ({\color{blue} \varphi^{\vee}}{\color{blue} \eta} + {\color{blue}\eta^{\vee}} {L}_v {\color{blue} \varphi}   \right ) ~.
    \end{equation}
% that satisfies the equivariant master equation {\color{red} cite}. In the above formulas tilda's stand for the fields without zero modes
%  and    now for new fields we introduce the fields without zero modes $({\color{blue} \tilde\varphi}, {\color{blue} \tilde{\varphi}^{\vee}},
%     {\color{blue}\tilde{\eta}}, {\color{blue} \tilde{\eta}^{\vee}} )$. The zero modes for these new fields decouple from the calculations and 
%      they can be ignored for further consideration.  
The new fields contain zero modes that must be decoupled from fluctuations, as we did for the UV fields. It is easy to see that zero modes decouple and so they can be ignored in the following. The extra term in the action just modifies the quadratic term (\ref{modified-SUV}) in UV fields
\begin{equation}\label{modified-SUV-extra}
 S_0'=S_0+\int \left ({\color{blue} \varphi^{\vee}}{\color{blue} \eta} + {\color{blue}\eta^{\vee}} {L}_v {\color{blue} \varphi}   \right ) 
\end{equation}
The following gauge fixing fermion (with ${L}_v$ invariant metric)
     \begin{equation}
      \Psi = \int {\color{blue} \tilde{\varphi}} (d\star {\color{blue}\tilde{\phi}_1})
     \end{equation} 
defines the following ${\mathcal L}_{UV}$
     \begin{equation}
      {\color{blue} \tilde{c}_4}=0~,~~~~{\color{blue}\tilde{\phi}_2^-}=0~,~~~{\color{blue}\tilde{\eta}^{\vee}}=0~,~~~
      {\color{blue}\tilde{c}_3} = \star d {\color{blue}\tilde{\varphi}}~,~~~~{\color{blue}\tilde{\varphi}^{\vee}} = d\star {\color{blue}\tilde{\phi}_1}~. 
     \end{equation}
Evaluating the quadratic part (\ref{modified-SUV-extra}) of the action we obtain
    \begin{equation}
S_0'|_{{\mathcal L}_{UV}}  =   \int \left ( {\color{blue} \tilde{\phi}_0} d \star d 
     {\color{blue}\tilde{\varphi}} + {\color{blue}\tilde{\phi}_1} d {\color{blue} \tilde{c}_2^-} +  d \star {\color{blue}\tilde{\phi}_1} {\color{blue}\tilde{\eta}}  \right  )\;.
     \end{equation}
      Introducing the inner product  $(~,~)$ as in (\ref{inner-product}) we can rewrite it 
%       as follows
%   \begin{equation}
%     - ({\color{blue}\tilde{\phi}_0}, \Delta {\color{blue}\tilde{\varphi}} ) - 
%     \frac{1}{2} ({\color{blue}\tilde{\phi}_1}, \star d {\color{blue}\tilde{c}_2^-}) - \frac{1}{2} ({\color{blue}\tilde{c}_2^-}, \pi^- d {\color{blue}\tilde{\phi}_1}) 
%     + \frac{1}{2} ({\color{blue}\tilde{\phi}_1}, d{\color{blue}\tilde{\eta}}) + \frac{1}{2} ({\color{blue}\tilde{\eta}}, d^\dagger {\color{blue}\tilde{\phi}_1})
%   \end{equation}
%   or  
  in the following matrix form
  \begin{equation}\label{quadratic_action}
  S_0'|_{{\mathcal L}_{UV}} = \frac{1}{2} (x,Dx)
\end{equation}
where $x=({\color{blue}\tilde{\phi}_0},{\color{blue}\tilde{\varphi}}, {\color{blue}\tilde{\phi}_1},{\color{blue}\tilde{c}_2^-},{\color{blue}\tilde{\eta}})\in V\equiv (1-P) \left ( \Omega^0 \oplus \Omega^0 \oplus \Omega^1 \oplus \Omega^{2-} \oplus \Omega^0 \right )$ and $D:V\rightarrow V$ is defined as
     \begin{equation}\label{differential_operator}
   D=   \left( 
\begin{array}{ccccc}
    0 &  -\Delta  & 0  & 0 & 0\\
  -\Delta  & 0 & 0  & 0 & 0\\
   0 & 0 & 0 &  - d^\dagger  & d  \\
  0 & 0 & -\pi^- d  & 0 & 0 \\
  0 & 0 & d^\dagger  & 0& 0 
\end{array}
\right)
\end{equation}
The differential operator $D$ is 
%defined on
%        $$(1-P) \left ( \Omega^0 \oplus \Omega^0 \oplus \Omega^1 \oplus \Omega^{2-} \oplus \Omega^0 \right ) $$ 
%     and 
by construction self-adjoint and invertible, its inverse being as follows 
       \begin{equation}\label{D-inverse}
D^{-1} = \left( 
\begin{array}{ccccc}
    0 &  -\Delta^{-1}  & 0  & 0 & 0\\
  -\Delta^{-1}  & 0 & 0  & 0 & 0\\
   0 & 0 & 0 &  - 2\Delta^{-1} d^\dagger  & \Delta^{-1} d  \\
  0 & 0 & -2 \pi^- \Delta^{-1} d  & 0 & 0 \\
  0 & 0 &  \Delta^{-1} d^\dagger  & 0& 0 
\end{array}
\right)\;.
\end{equation}
In order to check it, one has to use the properties (\ref{Dleta-pi-property}) and the explicit definition of $\Delta$ on zero forms. 
  
Next the restriction of the linear term $S_1$ in (\ref{mixed-first}) to ${\mathcal L}_{UV}$ gives 
\begin{equation}
       S_1|_{{\mathcal L}_{UV}}  = \int \left ( {\color{blue} \tilde{\phi}_0} {\color{red} \phi_4} - \star d {\color{blue}\tilde{\varphi}} ( \iota_v {\color{red}\phi_2^+}  + \iota_v {\color{red}\sigma_2^-}) + {\color{blue}\tilde{\phi}_1} ({\color{red}\phi_3} + d {\color{red} c_2^+} + \iota_v {\color{red}h_4})  + {\color{blue}\tilde{c}_2^-} \iota_v 
       {\color{red}\phi_3} \right )~,
\end{equation}
that can be written as
%  Using the inner product on differential forms we can rewrite it as follows
%  \begin{equation}
%    S_1 = ({\color{blue} \tilde{\phi}_0}, \star {\color{red} \phi_4}) - ({\color{blue}\tilde{\varphi}}, d^\dagger \iota_v ({\color{red} \phi_2^+}+{\color{red}\sigma_2^-}) ) 
%    - ({\color{blue}\tilde{\phi}_1}, \star {\color{red} \phi_3} - d^\dagger {\color{red} c_2^+} + \star \iota_v {\color{red} h_4})
%     - ({\color{blue}\tilde{c}_2^-}, \pi^- \iota_v {\color{red} \phi_3})~,
%  \end{equation}
%   where the fields with tilda's live in $(1-P)$ space. 
as
$$ S_1|_{{\mathcal L}_{UV}}= (x,a) $$
with $a\in V$ being
$$
   a = (1-P)\left (\star {\color{red} \phi_4},-d^\dagger \iota_v ({\color{red} \phi_2^+}+{\color{red}\sigma_2^-}), -\star {\color{red} \phi_3} + d^\dagger {\color{red} c_2^+} - \star \iota_v {\color{red} h_4},- \pi^-\iota_v {\color{red} \phi_3},0 \right ) \;.
$$
Combining these terms together we have the following structure
 \begin{eqnarray*}
(S_0'+S_1)_{{\mathcal L}_{UV}} &=& \frac{1}{2} (x, Dx) + (x,a) \cr
&=& \frac{1}{2} (x+ D^{-1}a , D (x + D^{-1} a)) - \frac{1}{2} (a, D^{-1} a) 
 \end{eqnarray*}
so that, after integrating out the fields in UV sector, we obtain
$$
  S_{\text{eff}} = \tilde{S}_{IR} -  \frac{1}{2} (a, D^{-1} a) ~. 
$$
with $\tilde{S}_{IR}$ defined in (\ref{IR-term-mixed}). Upon  integration we also obtain the field independent determinants which 
however depend on the metric.
  In our consideration we ignore those determinants, but for physical discussions (e.g., like metric dependence etc) they do play an important role. 
Since we are left with red fields, IR fields and UV zero modes, from now on we stop the colour marking of the fields.  After some computations one gets  
% $$      - \frac{1}{2} (a, D^{-1} a)  = -(\star \phi_4, K \iota_v (\phi_2^+ + \sigma_2^-)) +  2 (\star \phi_3- d^\dagger c_2^+ + \star 
%        \iota_v h_4, K
%        \pi^- \iota_v \phi_3)~,$$
% where $K= \Delta^{-1} d^\dagger$. Summarizing everything we obtain 
the following effective BV action for the IR fields
    \begin{eqnarray}\label{IR-eff-full}
  &  S_{\text{eq-eff}} &= \int \left( \phi_2^+ d   c_1 +   \phi_3 d   c_0 + \frac{1}{2} ( \phi_2^+)^2 + \sigma_0  \phi_4 + 
      \sigma_1  \phi_3++   \frac{1}{2} (\sigma_2^-)^2\right .\nonumber\\
& &  \phi_3 \iota_v  c_2^+ + \phi_4 \iota_v  c_1 +  \sigma_1 \iota_v  h_4 +   \sigma_2^- \iota_v   h_3
   +   \phi_3 \iota_v   h_2^- +  \phi_2^+ \iota_v  h_3 \\ 
 &&      \left . - \phi_4 K\iota_v (\phi_2^+ + \sigma_2^-) -2( \phi_3   +  \iota_v h_4 ) K\pi^- \iota_v \phi_3  -2 c_2^+ dK\pi^- \iota_v \phi_3
     \right ) \nonumber\;.
    \end{eqnarray}
By construction it satisfies the equivariant master equation (\ref{equiv_CME}) with 
   \begin{eqnarray}
    S^\text{eff}_{\Hat L_{v_a}}  = \int \left (\phi_4 L_v c_0 + \phi_3 L_v c_1 + \phi_2^+ L_v c_2^+  + \sigma_0 L_v h_4 + 
    \sigma_1 L_v h_3 + \sigma_2^- L_v h_2^-  \right ) \nonumber
      \end{eqnarray}

 By putting $v=0$ one recognizes the non equivariant effective action in (\ref{effective_action_nonequiv}) which coincides, up to zero modes, with the BV extension of abelian Yang-Mills theory. We can say then that (\ref{IR-eff-full}) is an equivariant extension of the abelian Yang-Mills BV theory.

We can read off the equivariant BV transformations for IR fields
\begin{eqnarray}\label{equivariant_differential_IR}
  &&\sfdeltaeq c_0 = \iota_v c_1 +\sigma_0 - K \iota_v \phi_2^+-K\iota_v\sigma_2^-\\
 & &\sfdeltaeq c_1 = d c_0 + \sigma_1 +\iota_v h_2^-+\iota_v c_2^+ 
 - 2 K (\pi^- \iota_v \phi_3)  
 \nonumber\\
 & & \phantom{\sfdeltaeq c_1 =}- 2 \iota_v (\pi^- K \phi_3)  + 2 
   \iota_v ( \pi^- d K c_2^+)-2\iota_v\pi^-K\iota_vh_4\nonumber\\
&  & \sfdeltaeq c_2^+ = \pi^+ \Big ( d c_1 + \phi_2^++ \iota_v h_3 - \iota_v (K \phi_4) \Big )\nonumber\\
  &&  \sfdeltaeq \phi_2^+ = \pi^+ \Big (  \iota_v \phi_3  -2 dK (\pi^- \iota_v \phi_3) \Big ) \nonumber \\
  && \sfdeltaeq \phi_3 = d\phi_2^+ + \iota_v \phi_4   \nonumber \\
 && \sfdeltaeq \phi_4 = d \phi_3 \nonumber
 \end{eqnarray}
 and for UV zero modes
\begin{eqnarray}\label{equivariant_differential_ZM}
 && \sfdeltaeq \sigma_0 = -2P\iota_vK\pi^-\iota_v\phi_3 + P \iota_v\sigma_1   \\
 &&\sfdeltaeq \sigma_1 = P\iota_{v}\phi_2^++P\iota_v\sigma_2^- \nonumber  \\
 && \sfdeltaeq \sigma_2^- =P \pi^- \iota_v\phi_3 \nonumber  \\
 && \sfdeltaeq h_2^- =\pi^-P\iota_v(-K\phi_4+h_3)+\sigma_2^- \nonumber  \\
 && \sfdeltaeq h_3 = P( \phi_3+\iota_vh_4) \nonumber \\
 &&\sfdeltaeq h_4 = P \phi_4\ \nonumber
\end{eqnarray}
The reader, with a straightforward but tedious calculation, can check that $\sfdeltaeq$ satisfies
\begin{equation}
\label{effective_equivariance}
\sfdeltaeq^2 = - \hat{L}_v~.
\end{equation}

\begin{rmk}
The invariant metric defining the UV gauge fixing is arbitrary. In the computation of (\ref{equivariant_differential_IR}) we implicitly assumed that it is in the same conformal class of the metric defining the IR space (or equivalently the YM theory). There is then a family of equivariant extensions parametrized by invariant metrics that are conformally equivalent to the {%11 YM metric
metric used for the definition of the YM action}. From the general discussion of subsection \ref{eq_BV_push_fwd} it follows that they give equivalent extensions of abelian YM. 
\end{rmk}

There are terms in $\sfdeltaeq$ that are quadratic in the vector field $v$. Restoring the formal parameter $u$ of degree $2$, one can expand $\sfdeltaeq$ in powers of $u$ and get 
\begin{equation}\label{general_form_eq_diff}
\sfdeltaeq=\sfdelta + u \sfdelta_{-1} + u^2\sfdelta_{-3}~~~~.
\end{equation}
From (\ref{effective_equivariance}) they satisfy the following relations
  \begin{eqnarray}\label{new-Cartan}
  &&  \sfdelta^2=0~, \nonumber \\
   && \sfdelta \sfdelta_{-1} + \sfdelta_{-1} \sfdelta = L_v~, \nonumber \\
 &&   \sfdelta \sfdelta_{-3} + \sfdelta_{-3} \sfdelta + \sfdelta_{-1}^2 =0 ~,\\
&&   \sfdelta_{-1} \sfdelta_{-3} + \sfdelta_{-3} \sfdelta_{-1} =0 ~, \nonumber\\
  &&  \sfdelta_{-3}^2=0 \nonumber
  \end{eqnarray}
The first two lines are the standard Cartan relations that make us to intepret $\sfdelta_{-1}$ as the contraction operator; the third lines states that $\sfdelta_{-1}$ squares to zero only up to $\ad(\sfdelta)$. {%We can think (\ref{new-Cartan}) as a higher analog of the Cartan calculus. 
This is an instance of the dgla introduced in \cite{AS} to build differential models of equivariant cohomology and was used in the BV context in \cite{M}.}      

\begin{rmk}
Analogously to Remark \ref{enlarged_costello} we can read from (\ref{equivariant_differential_IR}) and (\ref{equivariant_differential_ZM}) the equivariant differential on the enlarged Costello complex (\ref{d-1-diagram}).
We can organize it as 
  \begin{equation}\label{eff-equiv-diff}
 d^{\rm{eff}}_u = d_{+1} + u d_{-1} + u^2 d_{-3}~,
 \end{equation}
 where $d_{+1}$ is the degree $1$-differential described in (\ref{d-1-diagram}). The operation $d_{-1}$ lowers the degree by $1$ and can be summarized by the following diagram
          $$\begin{tikzcd}
 & \Omega^0   &  \arrow[l, yshift=0.8ex, swap, "\iota_v"]  \Omega^1 &   \arrow[l, yshift=0.8ex, swap, "\iota_v (1+ 2 \pi^- dK)\oplus \iota_v"]  \Omega^{2+} \oplus H^{2-}     &   H^3   \arrow[l, yshift=0.8ex, swap, "\pi^+\iota_v \oplus \pi^- P \iota_v"] &   \arrow[l, yshift=0.8ex,  swap,  "P\iota_v"] H^4   \\
   H^0     &  \arrow[l, yshift=-0.8ex, "P\iota_v"]   H^1     & \arrow[l,  yshift=-0.8ex, "P\iota_v \oplus P \iota_v"] \arrow[ul, xshift= -1.0ex, "-(K\iota_v \oplus K\iota_v)" description]  \Omega^{2+}\oplus H^{2-}     &   \arrow[ul, xshift=-1.0ex, swap,  "-2(K\pi^- \iota_v + \iota_v \pi^- K)" description]    \Omega^3  \arrow[l, yshift=-0.8ex," \pi^+(\iota_v - 2 dK\pi^- \iota_v) \oplus \pi^- P\iota_v"]  &   \arrow[l, yshift=-0.8ex, "\iota_v"] \arrow[ul,  xshift= 1.0ex, swap,  "-(\pi^+ \iota_v K \oplus P\pi^- \iota_v K)" description]  \Omega^4   &
     \end{tikzcd}$$
  The last operation $d_{-3}$ (the terms quadratic in $v$) lowers the degree by $3$ and the only non-trivial 
    operations are given by the following diagram 
    $$\begin{tikzcd}
 & \Omega^0   &   \Omega^1  &    \Omega^{2+} \oplus H^{2-}    &   H^3    &  \arrow[bend right]{lll}[swap]{-2\iota_v \pi^- K \iota_v}  H^4   \\
   H^0     &   H^1     & \Omega^{2+}\oplus H^{2-}     &  \arrow[bend left]{lll}{-2P \iota_v K \pi^- \iota_v}  \Omega^3   &  \Omega^4   &
     \end{tikzcd}$$
By construction we know that
 \begin{equation}
 (d^{\rm{eff}}_u)^2 = (d_{+1} + u d_{-1} + u^2 d_{-3})^2 = u L_v~,
 \end{equation}  
so that we have the same generalized Cartan calculus as in (\ref{new-Cartan}). 
\end{rmk}

\begin{rmk}
Due to the dependence on the homotopy operator $K$, the equivariant operator $d^{\rm{eff}}_u$ is not a derivation. This can be checked on the Costello complex ignoring the contribution of zero modes. As a consequence, the effective action (\ref{IR-eff-full}) has non zero bracket with the cubic term $\int{\mathbf A}^3$ so that the equivariant extension of the non abelian case cannot be obtained by the sum of (\ref{IR-eff-full}) and the cubic term. The only possible approach is the computation of the BV-pushforward, which
 will involve all orders in perturbation theory. Due to degree considerations the general structure of $\sfdeltaeq $ is the same as in (\ref{general_form_eq_diff}), with $\sfdelta$ being the effective BV differential defined by the (non equivariant) BV pushforward. 
\end{rmk}

\section{Summary}\label{s:summary}

In this paper we considered two four dimensional BV theories: DW theory associated to BV manifold ${\mathcal M}_{DW}$ and YM theory based on 
 BV manifold  ${\mathcal M}_{IR}$ (see the text for the definitions). There exists a formal push-forward map from ${\mathcal M}_{DW}$  to  ${\mathcal M}_{IR}$
  and thus on  general ground we expect that the two theories are quasi-isomorphic at the level of observables. For example, a local simple observable in one theory can 
   be mapped into something complicated and non-local in another theory. 
   We do not know how much this formal statement is useful for actual calculations in one of the theories. 
    In this paper we have performed the BV push-forward map for the case of abelian theories, and in the case of equivariant theory we obtain  non-local 
     corrections to the BV YM action. 
       The equivariant extension of the BV formalism was originally introduced in \cite{equiv-BV}  
       and in this setting  upon an appropriate gauge fixing  the residual BRST symmetries should 
       realize  the equivariant differential on the space of fields
        (or some version of equivariant differential). Such residual  BRST symmetries may allow to perform the exact calculations
         for some theories and for some specific observables. 
       Thus   we hope that these new structures presented here 
       can be useful for the calculations in the same fashion as the equivariant considerations help 
      calculate the ordinary integrals of the closed differential forms. But this requires further study. 

Another interesting result of this work is the homological generalization of Cartan calculus with the explicit constructions in terms of non-local operators. 
 This structure requires further study and better understanding both at the level of the formal properties and at the level of concrete examples. 
  It would be interesting to see if any of the standard considerations for the equivariant cohomology and equivariant localization can be 
   generalized to this new setting. 

\appendix

\section{Hodge decomposition and relevant operators}\label{appendix_Hodge}

In this appendix we collect the relevant properties of the Hodge decomposition and the related differential 
 operators. All formulas are written for a 4d {%12 
 closed} manifold $\Sigma_4$.  Let $d$ be {%13
 the} de Rham operator. We pick up 
  the metric and introduce the inner product on differential forms 
\begin{equation}\label{inner-product}
 (\alpha, \beta) = \int \alpha \wedge \star \beta~,
\end{equation}
 where $\star$ is the Hodge star operator with the property
 \begin{equation}
   \star^2 \omega_r = (-1)^r \omega_r~, 
 \end{equation}
  where $\omega_r \in \Omega^r(\Sigma_4)$.  We define the adjoint operator $d^\dagger$ for 
   the de Rham differential
   \begin{equation}
     (\alpha, d\beta) = (d^\dagger \alpha, \beta)
   \end{equation}
    with the explicit definition given by
    \begin{equation}
     d^\dagger = - \star d \star~.
    \end{equation}
     Next we define the Laplace operator
     \begin{equation}
      \Delta = d d^\dagger + d^\dagger d~,
     \end{equation}
      which is a self-adjoint operator $\Delta^\dagger = \Delta$. The Hodge decomposition for $\Omega^r$
       corresponds to the decomposition into three orthogonal spaces
       \begin{equation}
        \Omega^r = ({\rm Im}~ d) \oplus ({\rm Im} ~d^\dagger) \oplus ({\rm Harm}^r)~,
       \end{equation}
       where $({\rm Harm}^r)=( {\rm ker}~\Delta)$ stands for harmonic forms which can be identified with the de Rham cohomology.
       We denote by $P$ the projector on $({\rm Harm}^r)$ and by $(1-P)$ the projector 
       to $({\rm Im}~ d) \oplus ({\rm Im} ~d^\dagger)$. Away from $({\rm Harm}^r)$ the Laplace operator is invertible, thus 
        whenever we write $\Delta^{-1}$ we assume that it is defined only on the $(1-P)$ space.  The operator  $\Delta^{-1}$ 
         is self-adjoint ($\Delta^{-1\dagger} = \Delta^{-1}$) and we have the following obvious properties
  $$ [d, \Delta]=0~,~~[d^\dagger, \Delta]=0~,~~ [d, \Delta^{-1}]=0~,~~ [d^\dagger, \Delta^{-1}]=0~,$$
   where the relations with $\Delta^{-1}$ understood only for the $(1-P)$ space. It is convenient to define the homotopy 
    operator
    \begin{equation}
      K= \int\limits_0^{\infty} dt~e^{-t\Delta} d^\dagger~,
    \end{equation} 
     which is defined everywhere and satisfies the property
     \begin{equation}
      K d + d K = 1 - P~.
     \end{equation} 
     If understood correctly $K$ can be expressed as follows
     \begin{equation}
      K = \Delta^{-1} d^\dagger~.
     \end{equation}
      On two forms we have the following projectors 
      \begin{equation}
      \pi^\pm = \frac{1}{2} (1 \pm \star)~,
      \end{equation}
      which splits
        $\Omega^2 = \Omega^{2+} \oplus \Omega^{2-}$ into self-dual and anti-self-dual two forms correspondingly. 
        We have the following useful properties
    \begin{equation}
  \pi^\pm \Delta = \Delta \pi^\pm~,~~~~~~ \pi^\pm \Delta^{-1} = \Delta^{-1} \pi^\pm
 \end{equation}
  and 
 \begin{equation}\label{Dleta-pi-property}
  \Delta \pi^\pm = 2 \pi^\pm d d^\dagger \pi^\pm ~.
 \end{equation}
    
Assume that $\Sigma_4$ admits the action of a vector field $v$ (typically coming from some group action).   
 On the differential forms $v$ acts via the Lie derivative 
 \begin{equation}
  { L}_v = d \iota_v + \iota_v d~,
 \end{equation}  
   where $\iota_v$ is contraction of $v$ with a differential form. We pick an invariant metric $g$ such that 
   ${\mathcal L}_v g=0$. In this case we have the following properties
   \begin{equation}
  [ {L}_v, d]=0~,~~~~
   [{L}_v, d^\dagger]=0~,~~~~[{L}_v, \Delta]=0~. 
  \end{equation}
 and thus ${L}_v$ preserves the subspaces in the corresponding Hodge decomposition.
 Moreover we have the following property
\bea
{L}_v^\dagger = - {L}_v~.
\eea

%\bibliography{Bibliography}

\begin{thebibliography}{6666}

\bibitem{AS} A. ~ Alekseev, P. ~ Severa, {\it Equivariant Cohomology and Current Algebras},  Confluentes Mathematici {\bf 4},2, 1250001 (2012), arxiv:1007.3118. 

\bibitem{AKSZ:geometry_of_BV}
M.~Alexandrov, M.~ Kontsevich,  A.~Schwarz and O.~Zaboronsky, 
{\it The Geometry of the Master Equation and Topological Quantum Field Theory},
Int.J.Mod.Phys., {\bf A12}, 1997, 1405-1430, hep-th/9502010

\bibitem{BV1}
  I.~A.~Batalin and G.~A.~Vilkovisky,
  {\it Relativistic S Matrix of Dynamical Systems with Boson and Fermion Constraints,}
  Phys.\ Lett.\  {\bf 69B} (1977) 309.
  
  \bibitem{BV2}
  I.~A.~Batalin and G.~A.~Vilkovisky,
  {\it Gauge Algebra and Quantization,}
  Phys.\ Lett.\  {\bf 102B} (1981) 27.

\bibitem{BV3}
I.~A.~Batalin and G.~A.~Vilkovisky,
 {\it Quantization of Gauge Theories with Linearly Dependent Generators,}
  Phys.\ Rev.\ D {\bf 28} (1983) 2567
   Erratum: [Phys.\ Rev.\ D {\bf 30} (1984) 508]

 \bibitem{equiv-BV}
F.~Bonechi, A.~S.~Cattaneo, J.~Qiu and M.~Zabzine,
{\it ``Equivariant Batalin\textendash{}Vilkovisky formalism,}
J. Geom. Phys. \textbf{154} (2020), 103720, 
 arXiv:1907.07995.
 
 \bibitem{BMZ}  
 F.~Bonechi, P.~Mn\"ev and M.~Zabzine,
{\it Finite dimensional AKSZ-BV theories,}
Lett. Math. Phys. \textbf{94} (2010), 197-228, 
 arXiv:0903.0995.
 
\bibitem{Ca-1}
 A. S. Cattaneo, P. Cotta-Ramusino, A. Gamba and M. Martellini, 
 {\it The Donaldson–Witten Invariants in Pure 4D-QCD with Order and Disorder
’t Hooft-like Operators,} Phys.\ Lett.\ B {\bf 355}, 245–254 (1995).
 
  \bibitem{CCFMRTZ}
  A.~S.~Cattaneo, P.~Cotta--Ramusino, F.~Fucito, M. ~Martellini, M.~Rinaldi, A.~Tanzini and M.~Zeni, 
  {\it Four-dimensional
Yang–Mills theory as a deformation of topological BF theory,}
 Commun.\ Math.\ Phys.\ {\bf 197}, 571–621 (1998).
 
  \bibitem{CattaneoMnevReshetikhin2018} A. S. Cattaneo, P. Mnev, N. Reshetikhin, 
  {\it Perturbative quantum gauge theories on manifolds with boundary,}  Commun.\ Math.\ Phys.\  {\bf 357.2}
   (2018) 631-730.

 \bibitem{CattaneoRossi2001} A. S. Cattaneo and C. A. Rossi, 
 {\it Higher-dimensional $BF$
theories in the Batalin--Vilkovisky formalism: the BV action and generalized
Wilson loops,}  \cmp{221}, 591\Ndash657 (2001).
 
 \bibitem{Costello} K.~Costello, 
 {\it Renormalization and effective field theory,}
 Mathematical Surveys and Monographs, \textbf{170}, 
  American Mathematical Society, Providence, RI, 2011. viii+251 pp. 
 
\bibitem{Ik} N.~Ikeda,  {\it Donaldson Invariants and Their Generalizations from AKSZ Topological Field Theories},  arXiv:1104.2100 [hep-th]

\bibitem{Khudaverdian:1989si}
  H.~M.~Khudaverdian,
  {\it Geometry of Superspace With Even and Odd Brackets},
  J.\ Math.\ Phys.\  {\bf 32} (1991) 1934.
  
  \bibitem{K-2}
  H. M. Khudaverdian, 
  {\it Semidensities on odd symplectic supermanifolds,}
   Commun.\ Math.\ Phys.\  {\bf 247}, 353-390 (2004)
   
\bibitem{M} A. Mikhailov {\it Insertion of vertex operators using BV formalism}, arxiv:2210.06745.

\bibitem{Mnev2008} P. Mn\"ev, ``Discrete $BF$ theory,'' \href{http://arxiv.org/abs/0809.1160}{arXiv:0809.1160}.


\bibitem{w88} E. Witten, {\it Topological quantum field theory}. Comm.Math.Phys. {\bf 117}, 353-386, (1988).

\end{thebibliography}

\end{document}